\documentclass[10pt,a4paper,twoside]{article}

\usepackage{indentfirst}            
\usepackage{graphicx}               
\usepackage{amsgen,amsfonts,amssymb,amsbsy} 
\usepackage{amsmath}%

\newenvironment{resum}{\begin{quote}\small}{\end{quote}}

\newcommand{\bfsf}[1]{\textsf{\textbf{#1}}}

\begin{document}

\thispagestyle{plain}       

\begin{center}


{\LARGE\bfsf{IMPLICATIONS  OF  A  TIME-VARYING\\[15pt]  FINE  STRUCTURE CONSTANT  }}

\bigskip


\textbf{Antonio Alfonso-Faus }


\emph{E.U.I.T. Aeron\'autica,  Spain}\/

\end{center}

\medskip


\begin{resum}
Much work has been done after the possibility of a fine structure constant being time-varying.
It has been taken as an indication of a time-varying speed of light. Here we prove that this
is not the case. We prove that the speed of light may or may not vary with time, independently
of the fine structure constant being constant or not. Time variations of the speed of light,
if present, have to be derived by some other means and not from the fine structure constant.
No implications based on the possible variations of the fine structure constant can be imposed
on the speed of light.

\end{resum}

\bigskip\noindent
PACS  numbers 04   04.60.-m

\bigskip


The strength of the electromagnetic interaction between photons and electrons is measured by
the fine structure constant, $\alpha= e^2/(4\pi\epsilon_0\hbar c)$. A time variation of this
constant has been recently reported as a slow increase over cosmological timescales [1-4].
Non-standard cosmological theories have been proposed invoking a varying speed of light [5-8]
or a varying electronic charge [9]. Using the entropy of a black hole [10] theoretical reasons
have been proposed to favor a varying $c$ over a varying $e$. However, we have already proved
elsewhere [11] that the speed of light does not enter in the formulation of the fine structure
constant. Hence no time variations of the speed of light can be inferred from time variations
of the fine structure constant.

All constants of physics are gravitational, with a varying degree of importance, due to the
fact that gravitation is present in all the Universe, and nothing can escape that. The
Equivalence Principle, and therefore Local Lorentz Invariance (LLI) and/or Local Position
invariance (LPI) [12], need not necessarily be violated by space and/or time variations of
certain constants, like the fine structure constant, provided first principles are preserved,
as the Action Principle for example. Preservation of Local Lorentz Invariance (LLI) as well as
Local Position Invariance (LPI)  [12], imply that it is necessary and sufficient for this
preservation that the permittivity $\epsilon_0$ and permeability $\mu_0$ be both equal to
$1/c$ [11]. Using a TH $\epsilon\mu$ formalism [12] one can implement the Einstein's
Equivalence Principle and prove that a necessary and sufficient condition for both LLI and LPI
to be valid is given by
\begin{equation*}
\epsilon_0=\mu_0=\left(H_0/T_0\right)^{1/2}
\end{equation*}
for all events. Since the product $\epsilon_0\mu_0$ is equal to $c^{-2}$, then the above
relation implies
\begin{equation*}
  \epsilon_0=\mu_0=\dfrac 1 c
\end{equation*}

Hence, in these units we can rephrase Maxwell's equations [11] giving Coulomb's law as $F =
ce^2/r^2$ and for the fine structure constant
\begin{equation*}
 \alpha=\dfrac{e^2}{4\pi\hbar}
\end{equation*}
so that the speed of light does not explicitly enter in the constitution of the fine structure
constant. The arguments using the entropy of a black hole to favor a varying $c$ over a
varying [10] $e$ are then drastically changed, under LLI and LPI preservation, to favor a
varying $\hbar$ and/or $e$ instead, and not $c$.


\end{document}